\documentclass[structabstract]{aa}
\usepackage{graphicx}
\usepackage{txfonts}
\begin{document}

\title{Do novae have optically thick winds during outburst with large
deviations from spherical symmetry?}

\author{M. Friedjung\thanks{Deceased on October 22, 2011.}}

\mail{roberto.viotti@iasf-roma.inaf.it}

\institute{Institut d'Astrophysique de Paris -UMR 7095,
CNRS/Universit\'e Pierre et Marie Curie, 98 bis Boulevard Arago,
75014 Paris, France}

\date{Received , Accepted, Draft September 29, 2011}

\titlerunning{ Non spherical winds of novae}
\authorrunning{M. Friedjung}

\abstract
{The evidence for the presence of optically thick winds, produced by
classical novae after optical maximum, has been challenged
in recent papers.
In addition, signs of orbital phase dependent photometric variations,
sometimes seen quite early in the development of nova outbursts,
are hard to interpret in the framework of optically thick
envelopes and especially winds.}
{A general discussion for belief in the presence of
optically thick winds with increasing ejection velocities
during the early stages of novae after their explosion, must be given.
This has to be done in order to clarify ideas about novae
as well as to contribute in particular to the understanding of the
behaviour of novae V1500~Cyg and V1493~Aql showing phase dependent
variations during very early decline after the outburst.}
{Possible ways of overcoming the apparent contradiction of phase
dependent variations through the production of deviations
from spherical symmetry of the winds, are looked at and
order of magnitude estimates are made for different theoretical
scenarios, which might produce such deviations.}
{It is found that large deviations from spherical symmetry of the
optically thick winds in early phases after the explosion can easily
explain  the problem of variations. In particular, the presence of a
magnetic field might have had a non-negligible
effect on the wind of V1500~Cyg, while at the present there is
not enough information available concerning V1493 Aql.}
{Optically thick winds/envelopes are almost certainly present in the
early stages  after optical maximum of a nova, while it is difficult to
make pure Hubble flow models fit the observations of those stages.
New more detailed observational and theoretical work, in particular
including the effects of magnetic fields on the winds, is needed.}

\keywords{Stars: binaries: novae -- Stars: individual: \mbox{V1493 Aql}
 -- Stars: individual: \mbox{V1500 Cyg} -- Stars: winds}

\maketitle

\section{Introduction}

Observations of novae soon after maximum brightness in the optical,
appear to require the presence of expanding optically thick envelopes
and probably winds, hiding the central remnant (e.g. Friedjung
\cite {F96}, Short et al. \cite{Sho01}). An envelope consists
of circumstellar material, while a wind with continuous ejection from
a central star is a special case. In this situation one does not
usually expect to observe photometric variations depending on the
orbital phase of the binary in the very early stages after maximum.
However, in a few cases, phase dependent
orbital variations seen very early after the nova outburst,
appear to pose a challenge to any sort of optically thick envelope
theory, including in particular  those { including} optically
thick winds. Models based on instantaneous ejection
of a thick envelope with a linear gradient of velocity,  or
``Hubble flow'' models, might seem to better fit the observations,
provided that the envelope started to become optically thin
when the variations are observed. 
Here we first describe the two { types of model} and then
give in some detail reasons for believing that the optically thick envelopes
of classical novae { after optical maximum contain} winds,
whose velocities increase with time.

Following that discussion we make a detailed examination of two novae
showing photometric variations of the order of magnitude
of the orbital period, very soon after optical maximum.
Finally, we try to see how the variations might be produced
because of deviations of the winds from spherical symmetry.

\section
{Processes in classical nova outbursts}

Following their explosions, classical novae show complex phenomena,
whose interpretation is not obvious. Observations indicate the
presence of regions, producing line absorption and line emission
with different velocities, belonging to what are called the
$``pre$--$maximum''$, $``principal''$, ``$diffuse$--$enhanced$''~ and $``Orion''$
systems. Each system tends to produce P~Cygni profiles
with absorption components of many spectrum lines
having very similar Doppler shifts, as well as central emission.
Such profiles of different systems are often superposed.
Each P~Cygni profile can be understood as being produced by material
expanding with a well defined range of velocities,
{ and becoming visible at different times}.

In this situation, the interpretation of spectral line profiles requires
care. Classical work, using photographic spectra showed that the
$``pre-maximum''$ system absorption components, seen before optical
maximum,  had a velocity which { sometimes decreased} with time;
the somewhat higher velocity $``principal''$ system appeared around maximum,
while the former system disappeared soon after.
According to such work, the Doppler broadening of the emission
lines of the $principal$ system is that of the emission lines of the
ejected nebula seen in the late {\it `` nebular''}
stage,  though weak $``Orion''$ emission wings have sometimes been
seen in the that stage (Payne-Gaposchkin \cite{P57}).
The higher velocity $``diffuse$--$enhanced''$ system, usually
appears some days later than the $principal$ system. Lines from more
ionised atoms belong to the $Orion$ system, which often (but not always)
has a higher velocity than the $diffuse$--$enhanced$ system.
Hence, systems with higher velocities tend to appear later than those
with lower velocities.
Absorption components of different systems with
different ionisations, such as those of the $principal$ and
$Orion$ systems, are simultaneously visible over typical timescales of
weeks. The velocity distribution and geometry are quite hard to interpret.
This is why there have been over the years strong disagreements
between different researchers who study the processes of
classical novae in the early stages after their explosions. These
disagreements involve the geometry and velocity field of the ejected
envelopes: either ejection is almost instantaneous,
or optically thick (in the continuum) winds play a major role
in continued ejection.

The first point of view is discussed in a recent review by Shore
(\cite{Sho08}). The model he supports is one of
``instantaneous ejection type II'' or ``Hubble flow'',
where after an initial acceleration, a geometrically
thick envelope has a linear gradient of velocity, with velocity
increasing outwards. The highest velocity material has travelled
the furthest and there is little change in the velocities and
geometry during the further evolution of the envelope.
Note that models, involving a Hubble flow, have been quite successfully
applied to supernovae.
According to this model  { an inhomogeneous}
nova envelope has both an inner and an outer edge, while a
still expanded white dwarf, undergoing nuclear burning
{ could} still lie below the inner edge.
The optical thickness of the
envelope decreases with time and the inner, more slowly expanding
regions at greater optical depths, become  visible in later
stages of post maximum nova development.
Shore (\cite{Sho08}) explains by this model { the shapes of}
{ certain} emission line profiles.
The narrowing with time of the Doppler broadened
emission line profiles, seen sometimes after optical maximum,
might { in addition be considered as strong evidence for Hubble flow
because of the increasing visibilty of the more slowly expanding
material with time. However Hubble flow neither explains the emission line
widening, often observed in earlier stages after maximum
(see Table 1), nor the rounded profiles typical of winds, seen in
 the earlier stages of for instance \object{V603 Aql} (Payne-Gaposhkin
\cite {P57}) and \object{V4169 Sgr} (Scott et al. \cite{Sc95})}.

\section {Evidence for the presence of optically thick winds
during outburst}

{ Models adding opticaly thick winds} are more elaborate, as they must
take account of the complex phenomena revealed by
the optical and ultraviolet observations of classical novae.
As discussed in early papers by McLaughlin (\cite{Mc47},
\cite{Mc65}), the outwards expansion velocities are higher in
deeper layers, which are nearer the centre of the exploding star
soon after optical maximum. More recently Seitter
(\cite{Se90}) came to similar conclusions.
McLaughlin explained the greater excitation/ionisation of the
high velocity systems, by they being due to material in the more central
regions of the envelope nearer the source of ionising radiation.
{ From UV observations of nova \object{V1974 Cyg}}
Cassatella et al. (\cite{Ca04}) found that,
when the velocities of absorption lines belonging to the same system
were compared, the velocities of lines of the more ionised atoms were
somewhat larger (70 km s$^{-1}$ between the \ion{Fe}{iii} ~lines
and the \ion{Fe}{ii} lines belonging to the $principal$ system).

There are other more precise reasons of why it is
necessary to believe in the presence of optically thick envelopes
with increasing ejection velocities  during the early stages after
the explosions of classical novae. In the two papers of McLaughlin,
just quoted, that conclusion is drawn when line emission and
absorption from different systems are superposed.
For instance, in the spectrum of \object{DQ Her} the $principal$
system absorption of
\ion{Sc}{ii} 4247 \AA~remained strong and sharp in the longwards wing
of the $diffuse$--$enhanced$ emission of \ion{Fe}{ii} 4233 \AA.
Moreover, that slowly developing nova showed
partial or complete obliteration of the $diffuse$--$enhanced$
absorption by overlying $principal$ emission. The strong
$diffuse$--$enhanced$ \ion{Si}{ii} multiplet (19) 6371 \AA~absorption was
obliterated in later February 1935 by [\ion{O}{i}] emission, unlike the
\ion{Si}{ii} 6347 \AA~line belonging to the same multiplet.
A similar argument was given by him for the obliteration of
the strongest multiplet (19) line of \ion{Ti}{ii} at 4395 \AA~in
the spectrum of that nova.
McLaughlin gives the same argument for the
$Orion$ system absorption of multiplet (1) \ion{N}{iii}~lines in the
spectrum of the fast nova \object{V603 Aql}. When one of
these lines was absent, it had shifted into coincidence with an
emission  maximum of \ion{O}{ii}~of the $principal$ system.

What is decisive is the result of combining ultraviolet continuum
observations with information derived from absorption line radial
velocities. Cassatella et al. (\cite{Ca02}) found that the
ultraviolet spectra of classical novae generally show a change in the continuum
colour temperature with decreasing optical brightness after optical maximum,
this being also shown by the appearance as the nova fades in the optical,
of lines of atoms with increasing ionisation (Cassatella et al. \cite{Ca05}).
The measurements in 20 \AA ~wide UV bands at 2885 \AA~ and 1455 \AA ,
that are free of strong line emission and absorption, indicate
a rising colour temperature soon after optical maximum, which can later fall,
possibly because of a contribution of the Balmer continuum.
The rising colour temperatures  are suggestive of a shrinking photosphere.
Such a conclusion is also reached just assuming the
bolometric luminosity constant during the fading of the optical continuum.
The photosphere, which according to Cassatella et al. (\cite{Ca04})
lasts about t$_3$, the time to fade 3 magnitudes, may have near
maximum a size not much smaller than the whole envelope, because the
beginning of the explosion is normally only a few days before maximum.
For example it was only about 5 days before maximum for \object{V1974 Cyg}
(Cassatella et al. \cite{Ca04}).
Soon after the envelope is much larger than the photosphere.
A detailed study by Friedjung (\cite{F87a}) of \object{FH Ser}, fitting
empirically the
energy distribution from the ultraviolet to the near infrared with a Planck
distribution,
indicated that the photosphere had shrunk by a factor of more than 5 over
several weeks.
Non--LTE modelling of the spectrum of \object{V1974 Cyg}
by Short et al. (\cite{Sho01}) indicated also a photospheric shrinkage
by about a factor of 3, the photosphere being there defined as
where the continuum optical depth at 5000 \AA~was equal to unity.
In the case of a Hubble flow, as already stated, more slowly expanding
material should have become more visible with time, as the optical
thickness of each layer of the envelope decreased.
But in fact \object{V1974 Cyg} showed an increase with time of
the velocity of the $principal$ system absorption, as usually occurs
for classical novae.
The same comment is also true for the $diffuse$--$enhanced$ absorption
lines of \object{V1974 Cyg}.

The behaviour of nova emission lines is less clear--cut
(Table 1), but narrowing was seen when the width was only measured after
the appearance of the $diffuse$--$enhanced$ system as for \object{V603 Aql},
\object{V 382 Vel} and \object{1974 Cyg}. Their emission line narrowing
can then be associated with decreasing emission from the
$diffuse$--$enhanced$ system compared with that of the $principal$ system, which
can be both due to a decreasing wind mass loss rate and increasing
ionisation with time of $principal$ system material.
\object{V1500 Cyg} is particularly instructive as the appearance of
$diffuse$--$enhanced$ absorption (Duerbeck and Wolf  \cite{DW77}) was
associated with rapid line widening  (Friedjung et al. \cite{Fr99}),
followed by slower line narrowing (Boyarchuk et al. \cite{Bo76}).

{ Therefore, after the disappearance of the $pre$--$maximum$ system,
observations appear to require the presence of an optically thick wind}, 
whose velocity increases with time, { formed} inside another region 
producing the $principal$ system. 
{ The optical thickness of the latter becomes small after optical maximum,
so allowing higher velocity deeper levels to become visible}. 
The higher velocity regions of the wind, in deeper layers, should
be then ejected later as long as this wind lasts. The ejection of
the $diffuse$--$enhanced$ and $Orion$ systems, appear
parts of the same physical process, that is wind production,
directly related to the photospheric radius and the mass ejection rate
at a particular time. Actually, using photospheric radii derived
from Zanstra--type temperatures, Friedjung (\cite{Fr66}) found that the
$diffuse$--$enhanced$ system absorption component velocities,
of \object{V603 Aql}, \object{RR Pic} and \object{DQ Her}, which had
post--optical maximum oscillations in their light curves,
corresponded to what would be expected for the varying $Orion$
system absorption component velocity, when the decreasing
photospheric radius was very large.
That suggests that the material producing the
lower velocity $diffuse$--$enhanced$ absorption, should be ejected earlier, (i.e.
when the photosphere is larger), than that producing the $Orion$ system.
Let us also consider that the disappearance of the $diffuse$--$enhanced$
system absorption has
been explained by the collision of a faster moving inner discrete shell
with that of the $principal$ system;
such an explanation was also recently suggested by Cassatella et al.
(\cite{Ca04}) for \object{V1974 Cyg}.
However, what may be involved could rather be
the collision of an absorbing cloud of the $diffuse$--$enhanced$ system
in the line of sight.

Let us note that a short lived Hubble flow {\it before optical
maximum  during the ``fireball'' stage}, can 
explain the properties of the $pre$-$maximum$ system. The velocity of its
absorption components { sometimes decreases with time}; a very large
effect of this kind was observed for \object{DQ Her}.
Perhaps velocity decreases of that system would be observed for all
novae, if they were observed early enough. The always larger velocity of
the $principal$ system absorption component shown by Payne-Gaposchkin
(\cite{P57}) for classical novae after maximum,
suggests that the $principal$ system material sweeps up all those parts
of the $pre-maximum$ system, which have a lower velocity. Friedjung
(\cite{F87b}) indeed suggested that the $principal$ system is formed
by the sweeping up of $pre$--$maximum$ material, by the faster optically
thick wind. Difficulties with this explanation were however found by
Cassatella et al. (\cite{Ca04}), in their analysis of
\object{V1974 Cyg}, so the situation appears to be less simple.

X--ray observations also suggest the presence of a wind, though they
do not indicate whether the wind is optically thick in the optical.
A shock front may be expected to be formed between a wind and the
region of formation of the principal system, with the production of
hot plasma and X--rays. A general review of X--ray observations has
been given by Krautter (\cite {Kr08}). Some of the observed X--rays
are soft and are understood as being emitted by a hot white dwarf
remnant in later stages of post optical maximum development. 
Harder X--rays with energies higher 
than $\sim$1~keV are most probably produced in a hot shocked plasma.
O'Brien et al. (\cite{Ob94}) performed theoretical calculations,
involving collisions between material ejected at different velocities,
with a fast wind blowing out the confining slow wind. They gave an
explanation for the detected X--rays of \object{838 Her}, though their
assumed constant mass loss rate of 4.74 10$^{-6}$~M$_{\odot}$~yr$^{-1}$
seems not realistic. Many other observers have
also detected hard X--ray emission such as Balman et al. (\cite{Ba98})
{ who derived a} maximum luminosity of 0.8--2.0~10$^{34}$ erg~s$^{-1}$
for \object{V1974 Cyg}. Mukai and Ishida (\cite {Mu01}) as well as
Orio et al. ({\cite{Or01}) studied X--ray emission of \object{V382 Vel}.
Tsujimoto et al. (\cite{Ts09} found for \object{V458 Vul}  
that a single temperature (0.64~KeV) hot plasma with a luminosity 
of 6~10$^{34}$~erg~s$^{-1}$ in the
0.3--3~Kev band and an emission measure of 7~10$^{57}$~cm$^{-3}$,
due to the collision between a wind and slower moving outer material, 
may well explain the observations.  
Results of later observations of X--ray emission
with the SWIFT satellite are given in recent papers, such as those by
Page et al. (\cite{Pa09}) of \object{V598 Pup} and by Ness et al.
(\cite{Ne09}) of \object{V2491 Cyg}. Page et al. (\cite{Pa09}) explain 
the observed late hard X--ray emission of
\object{V598 Pup} between 147 and 255 days after the outburst by
collisions between material having differential motion between 400 and
800~km~s$^{-1}$. 
However, Russel et al. (\cite{Ru07}) found that the
infrared lines had a HWFM of 2000~km~s$^{-1}$, so the X--ray
production may have there involved more complicated physics.
These X--ray luminosities can be compared with those of normal O
stars, where X--rays are thought to be produced by collisions inside their
winds. Naz\'e (\cite{Na10}) found, from XMN-Newton observations, that
the mean log ratio of O star 0.5--10~KeV X--ray emission to the bolometric
luminosity was $-6.45 \pm 0.51$. Using the basic stellar parameters
of Drilling and Landolt (\cite{DL00}), this result for an O5 main sequence
star would correspond to a flux of 1.2~10$^{33}$~erg~s$^{-1}$,
 much smaller than the above mentioned X--ray fluxes of novae. 
 
Finally let us note { following the X--ray discussion, that}
collisions { involving the wind and slower moving outer material,
like in the case of spherical symmetry studied by O'Brien et al
(\cite{Ob94}), have not been sufficiently examined up to the present time,
to give predictions for novae of different classes. In addition
instabilities can occur  with the wind  breaking up} the $principal$
system formation region.
In that way holes and separate clouds may be created. Some of the
faster unshocked cooler material might therefore overtake the lower
velocity clouds and then settle down into a Hubble flow after the
disappearance of the wind. In addition material belonging to the
fastest regions of the pre--maximum system, which were not swept up
by the principal system material, could also still be present, so
contributing to a Hubble flow. { Such a situation is difficult to
calculate, making numerical estimates unreliable.} Indications of that
sort of structure of the ejected nebula of \object{V1974 Cyg} two years
after outburst, with faster moving material at the outside, were
given by Panagia (\cite{Pa02}). In this connection we can
also mention the work of Ederoclite et al. (\cite{Ed06}) on
\object{V5114 Sgr}, who found that the mass
of H$\alpha$ emitting clumps as well as their filling factor
decreased with time; that can be explained if the clumps dissolved
slowly into lower density more massive material, producing less
emission.

We can conclude that though much more work remains to be done to
understand all the processes involved and to refine the
observational evidence, models including optically thick winds 
have a much larger power to
explain many different sorts of observation. Nevertheless,
at both very early and quite late times, the
kinematics of certain regions can be one of a Hubble flow.
{ A} nova wind, unlike the more common O star winds, can in principle
be accelerated by radiation pressure at large continuum optical depths,
as for instance discussed by
Friedjung (\cite{Fr66}) and Kato and Hachisu (\cite{KH94}).

\begin{table}
\caption{Temporal behaviour of typical novae}
\label{tab:temp}
\begin{flushleft}
\begin{tabular}{l c c l }
\hline\hline 
\noalign{\smallskip}
 Nova & t$_3$(days) & line width & remarks \\
\noalign{\smallskip}
\hline
\noalign{\smallskip}
\object{V1500 Cyg}& 3.6& w-n      &  (a) (1) (2) \\
\object{V1493 Aql}&  7  &  n      & (b) (3) \\
\object{V603 Aql} &  8  &  n      &  (4) \\
\object{V458 Vul} &  9  &  n      &  (5) \\
\object{V1722 Aql}& 16  & w-n     &  (6) \\
\object{V2491 Cyg}& 21: & n-w     & (c) (7)\\
\object{V382 Vel} & 21  & n       &  (8)  \\
\object{V5114 Sgr}& 21  & w       &  (9)  \\
\object{V1974 Cyg}& 42  & n       & (10)  \\
\object{PW Vul}   & 97  & w       & (11)  \\
\object{RR Pic}   &150  &         & (d) (12)  \\
\object{HR Del}   &230  &         & (e) (13)  \\
\noalign{\smallskip}
\hline       
\end{tabular}
\end{flushleft}
{{\it Symbols:}~~$n$, $w$~general narrowing/widening of emission lines' width with time,~~
$n-w$ ~emission line narrowing before widening,~~
$w-n$ ~emission line widening before narrowing.

{\it Notes:} after maximum (a) very rapid widening of the emission lines,
(b) no absorption seen, (c) H$\alpha$ width, probably no diffuse--enhanced 
absorption system seen, (d) width stable when seen, (e) width
of each emission component of profile increasing.

(1) Friedjung et al. (\cite{Fr99}), (2) Boyarchuk et al. (\cite{Bo76}),
(3) Arkipova et al. (\cite{Ar02}), (4) Payne-Gaposchkin (\cite{P57}),
(5) Tarasova (\cite{Ta07}), (6) Munari et al. (\cite{MH10}),
(7)  Munari et al. (\cite{Mu11}), (8) Della Valle et al.  (\cite{De02}),
(9)  Ederoclite et al. (\cite{Ed06}), (10) Cassatella et al. (\cite{Ca04}),
(11) Andrillat and Houziaux (\cite{AH87}) ,
(12) Payne-Gaposchlkin (\cite{P57}),
(13) Hutchings (\cite{Hu70}) }
\end{table}

\section{Novae with periodic photometric variations
very early after the optical maximum}

Photometric variation with a period consistent with the orbital period
of the nova binary systems has been observed in a number of novae after the
optical maximum. Here we discuss the most significant cases of two fast
novae.

A classical case of such variations is \object{V1500 Cyg}, which showed
periodic photometric variations as early as 10 days after maximum and
perhaps even 5 days after maximum according to Rosino and Tempesti
(\cite{RT77}). The spectroscopic line profiles showed signs of even
earlier variations at about 3 days after maximum according to Hutchings
et al. (\cite{Hu78}). This nova is now understood as being a polar, that is
as containing a white dwarf with a very strong magnetic field. In this
framework, the decrease of period, understood as being the white dwarf's
rotation period (which is not far from the orbital period), can moreover
be explained. The periodic photometric variations, found by Campbell
(\cite{Cam76}), were interpreted by a model with spherical symmetry.
Hutchings et al. (\cite{Hu78}) proposed a rotating searchlight model,
with a polar axis inclined 50$^{\circ}$ to the line of sight. In later work
Horne and Schneider (\cite{HS89}) studied 1981 radial velocity and flux
variations. They concluded that the emission lines arose near the
secondary component of the binary. The binary inclination was according
to them equal to or more than 40$^{\circ}$.

It should be noted that the emission lines of highly ionised
atoms appeared later than the periodic variations of
\object{V1500 Cyg}, according to the observations of Rosino and
Tempesti (\cite{RT77}) and by Hutchings et al. (\cite{Hu78}). For
example \ion{He}{ii}, produced by the recombination of twice
ionised helium, was only detected two weeks after maximum, which is
explicable by supposing that radiation, able to doubly ionise helium,
at wavelengths where the optical thickness of the wind was large, was
absorbed at earlier dates, without needing to suppose deviations from
spherical symmetry.

It is necessary to emphasize that the proposed deviations from spherical
symmetry are those of the wind, which need not be those of the
{ massive outer} envelope.
The rapidly disappearing wind would appear to have much
less mass than { the latter}. Slavin et al. (\cite{Sl95})
found that the image of the nebular ejecta of \object{V1500 Cyg} was
nearly circular with randomly scattered enhancements; in addition they
proposed a correlation between the aspect ratio of the nebular ejecta
and the speed of development of novae after optical maximum, so that
faster developing novae like \object{V1500 Cyg} have more spherical
envelopes. However Downes and Duerbeck (\cite{Do00}) were doubtful about
the statistical significance of the correlation. \par \bigskip

The short period variations of \object{V1493 Aql} appear to be much
more difficult to explain in a framework of spherical symmetry.
Photometric variations with a period of 0.156 days were observed, the
first signs being seen only 5 days after maximum, as found by Dobrotka et
al (\cite{Do05}) and by Novak et al. (\cite{No99}). The light curve
moreover suggests the presence of eclipses. The development of this nova
after maximum was slower than that of \object{V1500 Cyg}, with a value of
t$_3$ of 7 days (see table 1).
The lines seen in the spectrum appear to be not those characteristic of a high
temperature small central object, needed for its luminosity to be near
the Eddington limit. It was for this reason that
Friedjung et al. (\cite{Fr06}) and
Dobrotka et al. (\cite{Do06}) conclude
that \object{V1493 Aql} was an object with a much lower
luminosity than that of a classical nova.
On the other hand,
most of the light curve of \object{V1493 Aql},
(except for the secondary peak which Hachisu and Kato (\cite{HK10})
suspected as being due to a magnetic field),
and the light curve of \object{V1500 Cyg}, fit the universal
decline law for the brightness of classical novae.
Let it be noted  that Bonifacio et al. (\cite{Bon00}) found
for \object{V1493 Aql}
a non classical time variation in $B$-$V$, with the nova being
apparently bluer near maximum.

We find however much less information for the less observed
\object{V1493 Aql} than for the well observed \object{V1500 Cyg}.
In fact the distance of the former of 4--5 kpc
(Arkipova et al. \cite{Ar02}, Hachisu and Kato {\cite{HK10})
is more than 4 times that of the latter.
Even much larger distances of 19 kpc and 26 kpc are estimated by
Bonifacio et al. (\cite{Bon00}) and Venturini et al. (\cite{Ve04}).

\section{
Deviations from spherical symmetry}

It is easy to see that a solution to the problem of variations over an
orbital period early in the development of a nova after its explosion
exists in principle, if the optically thick wind (or envelope)
has large departures from
spherical symmetry. In particular if the wind is considerably stronger
near the poles with a considerably larger optical depth than near the
equator, it might be possible to detect eclipses of the central binary.
In that case the inclination of the orbit should be large. Explanations
involving the presence of the companion star, rotation of the source
of the wind and/or magnetic fields can indeed be imagined. \par \bigskip

\noindent {\bf Presence of the companion star} \par

According to nova wind models, soon after maximum
the companion star should be revolving in deep layers
of the wind. One might expect, in a rather simplistic way,
that such effects would produce a spiral disturbance in the wind, as was
in fact suggested by Fabian and Pringle (\cite{FB77}). The bottom of the
spiral should rotate with the orbital period, while if angular
momentum is conserved in the wind, the top moves much more slowly,
producing only long timescale variations. As the mass flux of the
wind decreased, radiation from lower more rapidly rotating parts of
the spiral would escape, so the variations would become more easily
detectable. The period of such variations would decrease during the
mass flux decrease, that is rapidly in the case of rapidly developing novae
after maximum like the fast novae \object{V1500 Cyg} and \object{V1493 Aql}.
However, the observed variations are not of this nature.

According to the models of Kato and Hachisu (\cite{KH94}), the
``drag'' luminosity produced by the companion star's motion in the
outflow, should be small in most cases. However their calculations,
made only for a one dimensional model, are unsuitable for predicting
to what extent there should be differences between the wind in the
plane of the orbit and perpendicularly to that plane.
Better ``2.5D'' calculations (according to the authors)
of the effects of the underlying binary system were performed by Lloyd et al.
(\cite{Ll97}). They take account of the ``common envelope phase'',
when the binary is far below the photosphere. The frictional drag of
the companion star causes orbital energy and angular momentum to be
transferred to the ejecta, resulting in a highly anisotropic flow.
The calculations permit rotation about the symmetry axis, though the
flow is constrained to be axisymmetric. The ejection velocity
increases with time. Fast novae, with high ejection velocities,
produced envelopes which were more spherical, while very low initial
ejection velocity novae produced envelopes with density enhancements
along the polar axis. However the two novae described above,
with photometric variations early in their post optical maximum
development, were fast with high ejection velocities.
It is therefore not clear, to what extent such calculations are
relevant to the present problem.
\par \bigskip

\noindent {\bf Rotation}

Rotation has been invoked by other authors to explain asymmetries
of the ejected envelope. Scott (\cite{Sc00}) considered the cooling
of the equatorial regions relative to the polar regions of the
underlying white dwarf if it were rotating fast. In that case the
pressure of the layers, and the Fermi temperature, needed to break
electron degeneracy and produce a thermonuclear runaway, would be
higher at the poles. The rate of energy production is indeed
extremely sensitive to the temperature and material ejected early in
the outburst, can show departures from spherical symmetry because of
rotation. Other effects are however needed to shape the wind.

Scott (\cite{Sc00}) recalls that the critical point of a rotating
wind will be higher near the equator, where the terminal velocity
is larger. In addition the flow near the equator is hindered by a
``centrifugal barrier'' there. He quotes work by Ruggles and Bath
(\cite{RB79}), and by Kato and Hachisu (\cite{KH94}), who predicted
higher photospheric wind velocities, when the critical radius is
smaller, but makes no  calculations himself on winds. Lamers and
Cassinelli (\cite{LC99}) discuss rotating winds more generally, both
for winds driven by radiation pressure in the lines and for
outflowing disks around stars. In the case of line--driven winds
effects become important with a rotation velocity above half that of
breakup (Friend and Abbott \cite{FA86}).
As far as outflowing disks are concerned, there is good
evidence for them around Be and B[e] stars; the suggested
bi--stability mechanism for their production depending on the
optical thickness in the Lyman continuum is not relevant to the
present discussion of winds accelerated at a large continuum optical
depth.

However, Owocki (\cite{Ow04}) pointed out that previous calculations
such as those of Friend and Abbott (\cite{FA86}) gave incorrect
results, because they did not take into account gravity darkening due
to von Zeipel's theorem. Similar considerations apply to winds of
``normal'' hot stars driven by radiation pressure in spectral lines
and those driven by radiation pressure acting on opacity of the
continuum, such as in the case of porosity modulated winds driven
by a luminosity above the Thomson scattering Eddington limit,
studied by Shaviv (\cite{Sha01}).
Outwards ``centrifugal acceleration'' reduces the effective
gravity near the equator. Owocki (\cite{Ow04}) examines the
situation when the critical point is near the stellar surface. If
von Zeipel's theorem describes  gravity darkening, the stellar
radiation flux scales with the effective gravity, taking into
account the ``centrifugal force''. Then

$$ F(\theta) = K ~(1 - \Omega ~sin^2 \theta) $$

Here $F$ is the stellar surface radiation flux and $K$ is a constant.
$\Omega$, the ratio of centrifugal force to gravity at the equator
equals $V^2_{rot}R/GM$, in terms of the stellar rotation velocity
$V_{rot}$, radius R, and mass $M$. $\theta$ is the co-latitude. Then
the effective Eddington parameter, which equals the ratio of
radiative acceleration to effective gravity, is independent of
latitude. When the distribution of the opacities of lines in the
case of line driven winds in CAK theory follows a power law
the mass flux at co-latitude $\theta$ turns out to be

$$ {m'_{\theta}\over{m'_0}} = 1 - \Omega ~sin^2 \theta $$


The result is that the wind is weakest near the equator. The
same sort of calculation can be made for { the previously mentioned
porosity modulated winds.} Again assuming von Zeipel's
theorem, the same
expression is found for $m'_{\theta}/m'_0$, when the luminosity is far above
the Eddington limit (see Owocki \cite{Ow04}). It is clear that, as
also for other sorts opacity variation, such rotational effects will
only be large when the slowly moving layers  near the base of the
wind are not too far from rotational breakup. In the case of
\object{V1500 Cyg}, for which the information is available, the
rotation velocity of the white dwarf in quiescence of the order of a
few km s$^{-1}$ is far below the velocity of rotational breakup of
the order of several thousands of km s$^{-1}$. Ejection of a wind
from originally slowly moving outer layers of a white dwarf
during the development of a nova outburst, would make the situation
worse for any effect involving rotation, if the angular momentum of
the ejected material were conserved. \par \bigskip

\noindent {\bf Magnetic Fields} \par
Magnetic fields will have a major effect, if the magnetic pressure
is not much less than the pressure accelerating the wind. In
order to make very approximate order of magnitude estimates, which
can be indications about what are the conditions, {\it when magnetic
fields start to become important}, it is easiest at the present
stage of knowledge to refer to published calculations, assuming the
continuity equation for spherical symmetry. Such estimates will fail
when magnetic fields are stronger.

Kato and Hachisu (\cite{KH94}) made calculations for optically thick
winds, where acceleration by a locally super-Eddington luminosity
occurs above the critical point, due to a change of the theoretical
OPAL opacity of the flow. In lower layers acceleration is by gas
pressure and we can use their fig. 4 for a wind of a 1.0~$M_\odot$
white dwarf with critical points at 0.2~$R_\odot$ and
0.65~$R_\odot$, to approximately estimate the gradients of gas
pressure there.
The order of magnitude of the gradient of the gas
pressure accelerating the wind (${v d(\rho v)/dr}$) ($\rho$ being the
density, $v$ the wind velocity and $r$ the distance from the centre
of the wind), should be compared with the order of magnitude of
the gradient of the
pressure of the magnetic field $d/dr{{H^2}/({8\pi})}$, needed to produce
substantial deviations of the wind from spherical symmetry. When
the continuity equation for spherical symmetry is valid, $\rho v$
varies as $r^{-2}$ and the pressure gradient of the wind equals
$2v{\rho}_0 v_0(r_0/r)^2 1/r$ with $v_0$ and ${\rho}_0$ the values of $v$
and $\rho$ at a reference radius $r_0$, which we shall take as that
at the critical point. A dipole field varies at large distances from
the dipole as $r^{-3}$ and the gradient of the pressure of the
magnetic field equals $(d/dr({{H_o^2}({r_0/r})^6\over{8\pi}})$,
with $H_0$ equal to $H$ at $r$ = $r_0$.
We find that, for the two critical points of 0.2 and 0.65 $R_{\odot}$,
magnetic fields of about 10$^{4.5}$ and 10$^3$ Gauss are
required near the white dwarf surface
to have a substantial effect on a spherically symmetric wind at
$r$ = $r_0$.

No detailed models of the porous winds studied by Shaviv
(\cite{Sha01}) are available, for which the effect of magnetic fields
can be estimated. However Kato and Hachisu (\cite{KH05}) modelled the
Thomson scattering super-Eddington luminosity of \object{V1974 Cyg},
with an artificially reduced opacity, to take account of the
clumpiness of the envelope. The fairly similar results for
different reduction factors of the luminosity with a white dwarf
mass of 1.0~$M_{\odot}$ are shown in their figure 1. The average position
of the critical point is at 10$^{11}$~cm. The magnetic fields at r =
10$^{8.8}$~cm, needed to have a substantial effect in accelerating
the wind near the critical point,
are then of the order of 10$^6$ Gauss.

The white dwarf of \object{V1500 Cyg} has a strong magnetic field,
of the order of 25 Mega Gauss according to the estimates quoted by
Warner (\cite{W95}) in his table 6.8. Our  estimates of the effect of a
magnetic field, indicate that it might produce major deviations from
spherical symmetry of its wind during outburst. As we have seen,
modelling of the system by Hutchings et al. (\cite{Hu78}) and by Horne
and Schneider (\cite{HS89}) suggest a large inclination of the polar axis to
the line of sight of more than 40$^{\circ}$,
so the wind could be weak in the direction of the line of sight.

Much less can be said about \object{V1493 Aql}, which has a larger
distance. It is not in the Chandra X--ray source catalogue and
its problem will requires further studies.

\section{Conclusions}

Firstly, we have emphasized that optically thick winds
are almost certainly present in the early stages
of a nova after its optical maximum.
Conversely, it is difficult to make rival { pure} Hubble flow
models fit the observations of those stages.

The winds may have large deviations from spherical symmetry with a
maximum { mass loss} along the polar axis, while when the total mass
in the wind is small compared with the the total mass of the envelope,
most ejected mass can be concentrated at the same time in a
nearly spherical envelope. Such deviations may enable variations with
a period of the order of the orbital period to be detectable soon
after optical maximum.
This could be the case of \object{V1500 Cyg} whose wind could be
strongly affected by its magnetic field, but much less can
be now said about \object{V1493 Aql}.
The order of magnitude estimates given in the present paper
are intended
to stimulate more detailed theoretical and observational work,
in particular including the effects of magnetic fields,
could be crucial.

\begin{acknowledgements}
The author must thank Izumi Hachisu and Stan Owocki for useful
discussions, as well as an anonymous referee who made useful suggestions
on the first draft of this paper. Roberto Viotti improved the style.
\end{acknowledgements}

\noindent{\it Note in memoriam.} 
Dr. Michael Friedjung passed away on October 22, 2011 in Paris  
just after having completed the present article to which he devoted 
his remaining energies. This work is the last of a 
long series of articles started 45 years ago, 
devoted to the study of the physics of classical novae 
during their explosive evolution. Michael Friedjung has been  
a worldwide known expert of interacting binaries; in particular, 
besides novae, he promoted important studies on symbiotic stars. 
He has also developed a useful tool, the 
{\it self absorption curve method}, 
to investigate hot sources with a rich emission line spectrum, 
in particular of \ion{Fe}{ii}, that can be applied 
to a large variety of astrophysical objects.


\begin{thebibliography}{}
\bibitem[1987]{AH87} Andrillat, Y., Houziaux, L. 1987, A\&A Suppl, 67, 111
\bibitem[1993]{Aa93} Annuk, K, Kolka, I., Leedjarv, L, 1993, A\&A, 269, l5
\bibitem[2002]{Ar02} Arkipova, V.P., Burlak, M.A., Esippov, V.F., 2002, Astron.
Rep., 28, 100
\bibitem[1998]{Ba98} Balman, S., Krautter, J., \"Ogelman, H., 1998, ApJ 499,
395
\bibitem[2000]{Bon00} Bonifacio, P., Selvelli, P.L., Caffau, E, 2000, A\&A,
356, L53
\bibitem[1976]{Bo76} Boyarchuk, A. A., Galinka, T. S., Gershberg, R. E.,
Krasnobabtsev, V. I., Rachkovskaya, T. M., Shakhovskaya, N. I., 1976,
Sov. Astron. Letters, 2, 166
\bibitem[1976]{Cam76} Campbell, B., 1976, ApJ 207, L41
\bibitem[2002]{Ca02} Cassatella, A., Altamore, A., Gonz\'alez-Riestra, R.,
2002, A\&A, 384, 1023
\bibitem[2004]{Ca04} Cassatella, A., Lamers, H.J.G.l.M., Rossi, C.,
Gonz\'alez-Riestra, R., 2004, A\&A 420, 571
\bibitem[2005]{Ca05} Cassatella, A., Altamore, A., Gonz\'alez-Riestra, R.,
2005, A\&A, 439, 305
\bibitem[2002]{De02} Della Valle, M., Pasquini, L., Daou, D., Williams, R.E.,
2002, A\&A 390, 155
\bibitem[2005]{Do05} Dobrotka, A., Retter, A., Hric, L., Nov\'ak, R., 2005,
in ``The Astrophysics of Cataclysmic Variables and Related Objects'', ed. J.-M.
Hameury \& J.-P. Lasota., ASP Conf. Ser., 330, 363
\bibitem[2006] {Do06} Dobrotka, A., Friedjung, M., Retter, A., Hric, L., Nov\'k,
R., 2006, A\&A, 448, 1107
\bibitem[2000]{Do00} Downes, R.A., Duerbeck, H.W., 2000, ApJ 120, 2007
\bibitem[2000]{DL00} Drilling, J.S., Landolt, A.U, 2000 in ``Astrophysical
Quantities'', fourth edition, ed A.N. Cox, Springer and AIP Press, 381
\bibitem[1977]{DW77} Duerbeck, H.W, Wolf, B., 1977, A\&AS, 29, 297
\bibitem[2006]{Ed06} Ederoclite, A., Mason, E., Della Valle, M., et al., 
2006, A\&A, 459, 875
\bibitem[1977]{FB77} Fabian, A.S., Pringle, J.E., 1977, MNRAS 180, 749
\bibitem[1966]{Fr66} Friedjung, M., 1966, MNRAS, 132, 317
\bibitem[1987a]{F87a} Friedjung, M., 1987a, A\&A, 179, 164
\bibitem[1987b]{F87b} Friedjung, M., 1987b, A\&A, 180, 155
\bibitem[1996]{F96} Friedjung, M., 1996, Mem. Astr. Soc. It., 67, 281
\bibitem[1999]{Fr99} Friedjung, M., Mikolajewska, J., Mikoajewski, M., 1999,
A\&A, 348, 475
\bibitem[2006]{Fr06} Friedjung, M., Dobrotka, A., Retter, A., Hric, L., Nov\'ak,
A., 2006, Astrophys. Space, Sci, 304, 317
\bibitem[1986]{FA86} Friend, D.B, Abbott, D.C, 1986, ApJ, 311.701
\bibitem[2010]{HK10} Hachisu, I., Kato, M., 2010, ApJ., 709, 680
\bibitem[1989]{HS89} Horne, K., Schneider, D.P., 1989, ApJ., 343, 888
\bibitem[1970]{Hu70} Hutchings, J.B., 1970, Publ. Dom Astrophys. Obs, 13, 347
\bibitem[1978]{Hu78} Hutchings, J.B., Bernard, J.E., Margetish, L., 1978,
ApJ 224, 899
\bibitem[1994]{KH94} Kato. M., Hachisu, I., 1994, ApJ., 437, 802
\bibitem[2005]{KH05} Kato, M., Hachisu, I., 2005, ApJ., 633, L117
\bibitem[2008]{Kr08} Krautter, J., 2008, in ``Classical Novae'', secomd
edition, ed. A. Evans, M.F, Bode 232
\bibitem[1999]{LC99} Lamers, H.J.G.L.M., Cassinelli, J.P., 1999,
``Introduction to Stellar Winds'', Cambridge University Press
\bibitem[1997]{Ll97} Lloyd, H.M., O'Brien, T.J., Bode, M.F., 1997, MNRAS.,
284, 187
\bibitem[1947]{Mc47} McLaughlin, D.B., 1947, PASP, 59, 244
\bibitem[1965]{Mc65} McLaughlin, D.B., 1965 in ``Colloque International sur
les Novae, Supernovae, Novoides'', Centre Naional de la Recherche Scientifique,
in discussion 123 also published in AnAp, 1964, 494
\bibitem[2001]{Mu01} Mukai, K., Ishida, M., 2001, ApJ 551, 1024
\bibitem[2010]{MH10} Munari, U., Henden, A., Valisa, P., Dallaporta, S.,
Righetti, G.L., 2010, PASP., 122, 898
\bibitem[2011]{Mu11}  Munari, U, Siviero, A., Dallaporta, S., Cherini, G., Valisa, P.,
Tomasella, L., 2011, arXiv:1009.0822vl
\bibitem[2010]{Na10} Naz\'e, Y., 2010, A\&A, 506, 1055
\bibitem[2009]{Ne09} Ness. J-U., et al., 2009, AJ 137, 4160
\bibitem[1999]{No99} Nov\'ak, R., Retter, A., Lipkin.Y., 1999, IAUC, 7443
\bibitem[1994]{Ob94} O'Brien, T.J., Lloyd, H.M., Bode, M.F., 1994, MNRAS., 271,
155
\bibitem[2001]{Or01} Orio, M., Parmar, A., Benjamin, R., et al.,
2001, MNRAS, 326, L13
\bibitem[2004]{Ow04} Owocki, S., 2004, in Evolution of Massive Stars, Mass Loss
and Winds, ed. M. Heydari-Malayeri, Ph. Stee, J.-P. Zahn, EDP Sciences, 163
\bibitem[2009]{Pa09} Page, K.L., Osborne, J.P., Read, A.M., Evans, P.A., Ness,
J-U, Beardmore, A.P., Bode, M.fF, Schwarz, G.J., Starrfield, S., 2009, A\&A,
507, 923
\bibitem[2002]{Pa02} Panagia, N., 2002, in discussion, Mem. Soc. Astron. Ital,
73, 251
\bibitem[1957]{P57} Payne-Gaposchkin, C, 1957, The Galactic Novae,
North Holland
\bibitem[1977]{RT77} Rosino, L., Tempesti, P., 1977. Sov. Astron., 21, 291
\bibitem[1979]{RB79} Ruggles, C.L.N., Bath, G.T., 1979, A\&A,80, 97
\bibitem[2007]{Ru07} Russel, R.W., Rudy, R.J., Lynch, D.K., Woodward, C.E.
2007, IAU Circ 8901
\bibitem[1995]{Sc95} Scott, A,D. et al., 1995, A\&A. 296, 439
\bibitem[2000]{Sc00} Scott, A.D., MNRAS, 2000, 313, 775
\bibitem[1990]{Se90} Seitter, W.C., 1990, in ``Physics of Classical Novae'',
ed. A. Cassatella \& R.Viotti, Springer, 79
\bibitem[2001]{Sha01} Shaviv, N.. 2001, MNRAS, 326, 126
\bibitem[2008]{Sho08} Shore, S.N., 2008, in ``Classical Novae'', 2nd 
edition, ed A. Evans, M.F. Bode, Cambridge, 194
\bibitem[2001]{Sho01} Short, C.I., Hauschildt, P., Starrfield, S., Baron, E.,
2001, ApJ, 547, 1057
\bibitem[1995]{Sl95} Slavin, A.J., O'Brien, T.J., Dunlop, J.S,, 1995, MNRAS,
276, 353
\bibitem[2007]{Ta07} Tarasova, T.N., 2007, IBVS 5807
\bibitem[2009]{Ts09} Tsujimoto, M., Takei, D., Drake, J.J., et al., 
2009, PASJ 61, S69
\bibitem[2004]{Ve04} Venturini, C.C., Rudy, R.J., Lynch, D.K., Mazuk, S.,
Puetter, R.C., 2004, ApJ, 128, 405
\bibitem[1995]{W95} Warner, B., 1995, Cataclysmic Vaiable Stars, Cambridge
University Press


\end{thebibliography}
\end{document}